\documentclass[]{iopart}
\usepackage{graphicx,epsf, epsfig, amssymb}
\usepackage[usenames]{color}

\newcommand{\be}{\begin{equation}}
\newcommand{\ee}{\end{equation}}
\newcommand{\bel}[1]{\begin{equation}\label{#1}}
\newcommand{\ba}{\begin{eqnarray}}
\newcommand{\ea}{\end{eqnarray}}
\newcommand{\bal}[1]{\begin{eqnarray}\label{#1}}

\newcommand{\url}[1]{{\it #1}}

\begin{document}

\title[Compact Binaries for LIGO and Virgo]{Compact Binary Coalescences in the Band of Ground-based Gravitational-Wave Detectors}
\author{Ilya Mandel$\footnote[1]{email: ilyamandel@chgk.info}, \footnote[2]{NSF Astronomy and Astrophysics Postdoctoral Fellow}$}
\address{Department of Physics and Astronomy, Northwestern University, Evanston, IL}
\author{Richard O'Shaughnessy$\footnote[3]{email: oshaughn@gravity.psu.edu}$}
\address{Pennsylvania State University, University Park, PA}

\begin{abstract}

As the ground-based gravitational-wave telescopes LIGO, Virgo, and GEO 600 approach the era of first detections, we review the current knowledge of the coalescence rates and the mass and spin distributions of merging neutron-star and black-hole binaries.  We emphasize the bi-directional connection between gravitational-wave astronomy and conventional astrophysics.  Astrophysical input will make possible informed decisions about optimal detector configurations and search techniques.  Meanwhile, rate upper limits, detected merger rates, and the distribution of masses and spins measured by gravitational-wave searches will constrain astrophysical parameters through comparisons with astrophysical models.  Future developments necessary to the success of gravitational-wave astronomy are discussed.

\end{abstract}

\maketitle

\section{Introduction}

We are likely only a few years away from the first direct detections of gravitational waves from coalescing compact-object (CO) binaries.  The Laser Interferometer Gravitational-Wave Observatory (LIGO, http://www.ligo.caltech.edu, \cite{LIGO}) has completed two years of data taking at the designed sensitivity (the S5 run) in the fall of 2007. Simultaneously, the Virgo detector (http://www.virgo.infn.it, \cite{Virgo}) ended its VSR1 run. Other ground-based detectors in operation include GEO 600 \cite{GEO600}.  By 2014-2015, the Advanced versions of the LIGO and Virgo detectors are expected to operate at sensitivities $\sim 10$ -- $15$ times greater than Initial LIGO \cite{AdvLIGO}. These instruments will be able to detect gravitational waves (GWs)  from inspirals, mergers, and ringdowns associated with coalescences of compact binary systems composed of neutron stars (NSs) or black holes (BHs). Although estimates of the rates of such coalescences cover a rather wide range, Advanced LIGO may be able to detect tens or even hundreds of these events per year (see, e.g., O'Shaughnessy et al.~\cite{OShaughnessy:2009}).

In this paper, we review the mechanisms of formation of coalescing NS and BH binaries and summarize the predictions for the merger rates of NS-NS, NS-BH, and BH-BH systems.  For NS-NS systems these predictions can be made by extrapolating from the small sample of observed binary pulsars, while for other binary types they are based on population-synthesis models of isolated binary evolution.  We discuss the potential importance of the dynamical formation of BH-BH binaries in dense stellar environments, such as globular and nuclear star clusters, and the possible coalescenses involving intermediate-mass black holes (IMBHs).  We summarize the existing knowledge about the primary characteristics of NSs and BHs in binaries, namely, their masses and spins. 

We focus on the tight connection between (i) astrophysical observations and modeling and (ii) gravitational-wave searches and data analysis.  Input from astrophysics can inform decisions about which detector configurations should be chosen, which  searches are the most likely to yield interesting results, and which waveform families should be used for detection.  On the other hand, looking beyond the excitement of the first detection, the detected merger rate (or even non-trivial upper limits) will allow us to limit the space of allowed astrophysical models, thereby providing constraints on currently ill-determined astrophysical parameters.  Even more precise constraints will come from matching the observed parameters of the binaries, particularly their masses and spins, with the model predictions.
 
\section{Mechanisms of Formation and Rate Predictions \label{sec:rates}}

The standard formation channel for compact-object binaries in the field is isolated binary evolution.  In this case, a primordial binary of main-sequence stars evolves through several stages of mass transfer 
and possibly common-envelope evolution, while the binary components age and eventually undergo supernovae, occasionally leaving behind a binary composed of neutron stars or black holes (see \cite{2007PhR...442...75K} for a review of this process).  A circular binary will merge through radiation-reaction from gravitational-wave emission on a timescale of \cite{Peters:1964}
\bel{Tmerge}
T_{\rm merge} \approx 6 \times 10^{8} \frac{M_\odot^3}{m_1 m_2 (m_1+m_2)}
        \left(\frac{a}{10^{11}\ {\rm cm}}\right)^4\ {\rm yr},
\ee
where $m_1$ and $m_2$ are the component masses and $a$ is the orbital radius.  In this section, we describe two different ways to evaluate the merger rates for CO binaries: directly from observations for NS-NS systems and via population-synthesis modeling for other binary types.  We also discuss dynamical formation scenarios that could be significant for BH-BH binaries in dense stellar environments.

\subsection{Binary pulsars: observational constraints and extrapolated NS-NS rates \label{extrap}}

Among double-CO binaries, NS-NS binaries are the only ones for which we have direct observational evidence.  This evidence comes from observations of Galactic binary pulsars: double neutron-star systems where at least one of the NSs is a pulsar.  Approximately ten such systems have been detected so far.  By modeling the selection effects of the searches which detected these systems, it is possible to estimate the total rate of merging NS-NS binaries in the galaxy, whether or not they are detected as binary pulsars (the pioneering efforts to do this are described in \cite{Phinney:1991ei} and \cite{Narayan:1991}).  Five of the detected systems are expected to merge within a Hubble time through radiation reaction from the emission of gravitational waves, and therefore contribute to extrapolated merger rate 
estimates.  These systems are:  B1913+16 (the Hulse-Taylor pulsar), B1534+12, J0737-3039 (a double pulsar), J1756-2251, and J1906+0746; there is also a sixth merging system, located in a globular cluster, which is thought to be atypical and is not generally included in merger rate calculations.

Kim et al.~\cite{Kim:2003kkl} developed the statistical framework necessary to reconstruct the Galactic NS-NS binary population from observations, while rigorously estimating the uncertainties which arise due to the small number of observed systems.  Kalogera et al.~\cite{Kalogera:2004tn} reported the extrapolated rates from the first three systems of the five listed above.  They used a variety of models for the pulsar luminosity distribution, which plays a very significant role in determining the selection effects.  The distribution is described by two variables: the minimum pulsar luminosity $L_{\min}$ and the slope of the luminosity distribution power law $p$.  For the preferred pulsar luminosity distribution with $L_{\rm min}=0.3$ mJy kpc$^2$ and $p=2$ (see model 6 in Table 1 of \cite{Kalogera:2004tn}), the authors found a most likely rate of $83$ NS-NS mergers per Myr in the Galaxy; the fifth and 95th percentile statistical rate estimates were $16.9$ and $292.1$ per Myr, respectively.  

Although such extrapolation has significantly fewer free parameters than the population-synthesis models described below, the numbers should nevertheless be used with caution.  Different choices of the variables describing the pulsar luminosity distribution, which are all consistent with the observations, could shift the merger rates by an order of magnitude.  The fifth percentile for the model yielding the most pessimistic merger rates suggests only $1$ merger per Myr in the Galaxy, while rates as high as $817.5$ per Myr are possible at the 95th percentile for the most optimistic model.  There are several possibilities for systematic biases related to the imperfect knowledge of pulsar age and the beaming fraction \cite{OShaughnessyKim:2009}.  The implicit assumption that the observed pulsars form a good representation of the overall Galactic NS-NS population is difficult to verify.  Lastly, the small number of observed systems means that a single system can significantly change the rates predictions.   For example, according to \cite{Kim:2006}, the inclusion of the relatively short-lived system J1906+0746 could increase the rate estimates above by almost a factor of $2$.  

\subsection{Population-synthesis models for isolated binaries \label{pop-synth}}

In the absence of direct observations of NS-BH and BH-BH binaries, the only way to predict the rates of binary mergers is through population-synthesis modeling.  A number of different models has been applied over the past two decades (see, e.g., \cite{Grishchuk:2001, lrr-2006-6, 2007PhR...442...75K} and references therein).  Generally, these models rely on a number of ill-constrained astrophysical parameters.  It is therefore necessary to explore the full allowed parameter space in order to quantify the uncertainties in the merger rates predictions.  We follow the procedure described in \cite{OShaughnessy:2008}, since most other models do not attempt a thorough exploration of the full parameter space.

The StarTrack population-synthesis code used in \cite{OShaughnessy:2008} has seven free parameters that can significantly affect the model outcomes, as identified in \cite{Belczynski:2008}.  These include: the power-law index in the binary mass ratio; 3 parameters used to describe the supernova kick velocity distribution; the strength of the massive stellar wind; common-envelope efficiency; and the fractional mass loss during non-conservative mass transfer.  Flat priors are used on these parameters to sample the widest possible set of models.  However, several additional constraints are then applied to disallow those models (i.e., those combinations of the seven input parameters) that disagree with observations.  The five constraints applied in \cite{OShaughnessy:2008} come from the following observations: (i) merging Galactic NS-NS binaries (i.e., binary pulsars expected to merge in a Hubble time); (ii) wide NS-NS binaries (those not expected to merge in a Hubble time); (iii) white dwarf - NS binaries; (iv) observed Type II supernova rate; and (v) observed Type Ib/c supernova rate.  Only the first three observations provide non-trivial constraints.  The predicted Galactic rates for NS-NS coalescences range from $5$ to $300$ per Myr, with the most likely value at $30$ per Myr \cite{OShaughnessy:2008}.  For NS-BH binaries, the range is from $0.05$ to $100$ per Myr, with the most likely value at $3$ per Myr \cite{OShaughnessy:2008}.  And for BH-BH binaries, model predictions range from $0.01$ to $30$ per Myr, with a most likely value of $0.4$ per Myr \cite{2007PhR...442...75K}.

In addition to the ranges stated above, which reflect the impact of different choices of astrophysical model parameters, further uncertainties exist.  For instance, if hypercritical accretion during the common-envelope phase is disallowed, merger rates can drop significantly \cite{VossTauris:2003, Belczynski:2007}.  Many potential BH-BH progenitors may enter the common-envelope phase while the donor star is evolving through the Hertzsprung gap; according to \cite{Belczynski:2007}, such systems may merge directly, thus significantly reducing the rate of BH-BH mergers.  If kicks are preferentially aligned with pre-supernova spin, NS-NS coalescence rates could decrease by up to a factor of five \cite{Postnov:2008}.   Additionally, uncertainties in the star-forming conditions, particularly metallicity, can lead to systematic errors in the rates predictions at least at the $\sim 30$-$50$\% level \cite{OShaughnessy:2008sfr}.  Recent calculations using updated models of stellar winds and direct observations of the black-hole X-ray binary IC10 X-1 suggest that low-metallicity environments may be even more important \cite{BulikBelczynski:2009, Belczynski:2009}.  Meanwhile, all of the applied observational constraints suffer from the uncertainties and possible biases discussed in the previous subsection.  Nonetheless, the default rate predictions for NS-BH and BH-BH merger from all of the population-synthesis models mentioned above, as well as those in \cite{Nelemans:2003} and \cite{Dewi:2006}, fall within the uncertainty range listed in the previous paragraph, suggesting that predictions generally converge within the uncertainties.

The rates quoted in Sections \ref{extrap} and \ref{pop-synth} are computed for our Galaxy.  To convert them to rates for other galaxies, they can be scaled by the star formation rate, which is tracked by the blue-light luminosity for spiral galaxies like the Milky Way \cite{LIGOS3S4Galaxies}.  However, while this tracer is relatively accurate for nearby galaxies, it fails to account for mergers that are significantly delayed relative to star formation in older elliptical galaxies \cite{pacheco:2005}.  This omission is particularly significant for BH-BH mergers, and therefore the actual BH-BH merger rate in the volume to which ground-based detectors will be sensitive in the Advanced LIGO-Virgo era can be significantly higher than the contribution from spiral galaxies alone would indicate \cite{OShaughnessy:2009}.

\subsection{Dynamical formation scenarios}

Dynamical effects can make a significant contribution to CO binary mergers in dense stellar environments, particularly globular star clusters and nuclear clusters at the centers of galaxies.  There, two-body encounters can lead to direct captures, while three-body single-binary interaction and four-body binary-binary interactions can tighten hard binaries to a point where radiation-reaction from gravitational-wave emission can take over and drive the binary to coalescence.

There have been some suggestions regarding the contribution of dynamically-formed binaries in globular clusters to short hard gamma ray bursts (SGRBs) \cite{Hopman:2006, Guetta:2009}, which are believed to be powered by mergers of NS-NS or NS-BH binaries.  However, extrapolating from the rate of observed SGRBs, as attempted, for example, in \cite{Dietz:2009}, is highly uncertain, because of unknown selection effects such as SGRB beaming fractions \cite{nakar07}, which may be different for field and cluster SGRBs \cite{Grindlay:2006}; additionally, binary mergers involving NSs are not the only possible formation scenario for SGRBs \cite{Virgili:2009, Lee:2009}.  In fact, most simulations indicate that dynamical effects are not important for binaries containing neutron stars, as heavier black holes will segregate to the dense cores and substitute in for neutron stars in binaries \cite{Sadowski, Ivanova:2008}. In particular, Grindlay et al.~\cite{Grindlay:2006} estimate the NS-NS merger rate from all globular clusters in the Galaxy to be 40 per Gyr, a factor of $\sim 1000$ less than predictions for field mergers.

On the other hand, dynamical effects can make a very significant contribution to BH-BH merger rates, since black holes are the more massive and will sink to the central, dense regions of the clusters.  O'Leary et al.~\cite{OLeary:2007} and Sadowski et al.~\cite{Sadowski} both examined BH-BH binary formation in globular clusters.  Their estimates range over several orders of magnitude, from $\sim 10^{-4}$ to $\sim 1$ BH-BH mergers per globular cluster per Myr, due to different assumptions about the fraction of star formation that occurs in clusters, the fraction of clusters that may form black-hole subclusters, and whether these BH subclusters decouple or remain in thermal equilibrium with other stars in the cluster.  Similarly, Banerjee et al.~\cite{Banerjee:2010} find that 3-body encounters in intermediate-age (a few Gyr) massive ($\gtrsim 3\times10^4$ $M_\odot$) clusters lead to the formation of a few merging BH-BH binaries per cluster, some of which coalesce after ejection from the cluster. O'Leary et al.~\cite{OLeary:2008} argue that the presence of a massive black hole in galactic nuclear clusters can create a sufficiently dense cusp to allow as many as $15$ BH-BH binaries per Gyr to form in such a cluster through direct 2-body encounters.  These values may be based on very optimistic assumptions regarding the fraction of BHs in galactic nuclei and the number density of low-mass galaxies, but if such BH binaries did form, they could be easily distinguished due to their significant eccentricity, which is not expected for other stellar-mass binaries in the band of ground-based detectors.  Meanwhile, Miller and Lauburg~\cite{MillerLauburg:2008} calculate that a few $\times \ 0.1$ BH-BH mergers per Myr per galaxy should occur in nuclear clusters of small galaxies without massive black holes, primarily through three-body interactions.

If intermediate-mass black holes (IMBHs) with a mass around $100$ $M_\odot$ are generic in globular clusters (see \cite{MillerColbert:2004, Miller:2009} for reviews on IMBHs, and \cite{2009Natur.460...73F} for an announcement of a recently discovered ultra-luminous X-ray source that represents a possible IMBH detection), they can contribute to binary merger rates.  Since an IMBH will be the most massive object in the cluster, it will readily sink to the center and substitute into a binary with a compact-object companion.  The binary will then harden through three-body interactions and eventually merge via an intermediate-mass-ratio inspiral on timescales of less than one billion years \cite{Mandel:2008}.  If the binary fraction in a young dense cluster is $\gtrsim 10$\% and the core-collapse time is $\lesssim 3$ Myr, two IMBHs can form in a cluster, yielding an IMBH-IMBH merger \cite{Fregeau:2006}.  Such a merger is also possible through the merger of two distinct globular clusters, each containing an IMBH \cite{Amaro:2006imbh}.  While Advanced LIGO and Virgo will be sensitive to coalescences involving IMBHs in globular clusters, third-generation detectors like the Einstein Telescope, with improved low-frequency sensitivity \cite{Freise:2009}, could detect high-redshift mergers of IMBHs formed from population III stars which may be the seeds of today's massive black holes \cite{Sesana:2009ET, Gair:2009ET, Gair:2009ETrev}.  However, the present lack of knowledge about the existence and occupation fraction of IMBHs in globular clusters and about early structure formation makes very speculative all rates predictions for mergers involving IMBHs.

\section{Expected characteristics of binaries \label{sec:char}}

The astrophysical characteristics of merging binaries, namely, their mass and spin distributions, are of great interest to gravitational-wave astrophysicists.  However, these are not well understood.  As with merger rates, there are two ways to constrain the binary parameters: through direct observations, and through modeling.  As discussed above, direct observations are limited to a few systems for NS-NS binaries, and non-existent for NS-BH and BH-BH binaries.  And although BH mass and spin measurements exist from mass-transfering X-ray binaries, these do not necessarily have the same mass and spin distributions as BHs in compact-object binaries because of the particular evolutionary histories of double-CO  binaries.  Meanwhile, models are sensitive to assumptions and can differ significantly in their predictions.  We discuss what is presently known about mass and spin distributions for NSs and BHs below.

\subsection{NS and BH masses}

The masses of $\sim 50$ pulsars in binaries have been estimated to date  \cite{Lorimer:2008,psr-ATNF-2005}.   Until a decade ago, most observed binary pulsar masses were extremely consistent with a ``canonical'' value of $1.35 M_\odot$ \cite{Thorsett:1998uc}.  Since then, several significantly more massive pulsars have been found with much larger confirmed masses, such as J1903+0327 with a confirmed timing mass $1.74 M_\odot$ \cite{psr-discovery-J1903+0327}.  In globular clusters, there are four pulsars in binaries with weakly constrained orbital inclinations whose masses appear to be higher than ``canonical'', from $1.7 M_\odot$ for a pulsar in Terzan 5 up to $2.74\pm 0.22 M_\odot$ for J1748-2021B; see Figure 28 in \cite{Lorimer:2008}.

More than 40 black-hole X-ray binaries have been detected to date.  The masses of approximately 20 of them have been measured by observing the companion dynamics \cite{RemillardMcClintock:2006,  Orosz:2007}.  These masses vary between $\sim 4$ and $\sim 20$ solar masses.   However, because of the mass-transfer episodes involving common envelopes, which play a crucial role in the formation of NS-BH and BH-BH binaries, these masses do not necessarily reflect the typical BH masses in such binaries.

In the absence of direct evidence, masses can be predicted by population-synthesis models.   Most studies in the literature focus on predictions of the chirp mass, $M_c=(m_1 m_2)^{3/5} / (m_1+m_2)^{1/5}$, since this parameter determines the gravitational-wave amplitude.  The chirp mass is $\approx 0.87$ of the total mass for for equal-mass binaries.  It is typically found to be between $\sim 2$ and $\sim 5$ solar masses for NS-BH systems and between $\sim 5$ and $\sim 10$ solar masses for BH-BH systems \cite{BulikBelczynski:2003, OShaughnessy:2005, Belczynski:2007, OShaughnessy:2008, OShaughnessy:2009}.  However, mean chirp mass values vary by $\gtrsim 30$\% between these predictions depending on assumptions about the cutoff NS mass, accretion rate, etc.  The mass ratio is typically very poorly constrained in existing population-synthesis studies (see, e.g., Fig.~4 of \cite{OShaughnessy:2008}).  

Meanwhile, a number of channels have been suggested through which significantly more massive black-hole binaries could form.  Recent observations of IC10 X-1, a binary with a massive black-hole accreting from a Wolf-Rayet companion star, suggest that more massive BH-BH binaries can form through isolated binary evolution, with chirp masses $\sim 20\ M_\odot$ \cite{Bulik:2008}.  Dynamical formation channels will favor relatively massive BH-BH binary components, as the most massive black holes are likely to sink to the centers of clusters through mass segregation and substitute into binaries during three-body encounters.  Moreover, the BH merger products in such dense clusters can be reused if they are not ejected from the cluster due to recoil kicks, leading to higher-mass mergers in subsequent generations; chirp masses for BH-BH mergers in globular clusters can therefore range to $\sim 30\ M_\odot$~\cite{Sadowski}. Additionally, although stellar winds in high-metallicity environments may prevent the formation of massive black holes, mass loss through stellar winds would be much less significant in low-metallicity environments, allowing more massive black holes to form \cite{Belczynski:2009}.  Finally, IMBHs discussed above could range to hundreds or thousands of solar masses, although the more massive IMBHs will not be detectable with ground-based detectors.

\subsection{NS and BH spins}

The observed pulsar spin periods and assumptions about NS spindown rates place the NS spin periods at birth in the range $10-140$ ms \cite{Lorimer:2008}, which corresponds to dimensionless spins $a/M \lesssim 0.04$.  Meanwhile, the fastest known recycled pulsar in a double NS binary has a spin of $a/M=0.02$ \cite{Burgay:2003jj}.  Thus, the spins of neutron stars are likely to be irrelevant for gravitational-wave detection and parameter estimation.

On the other hand, direct measurements of BH spins in X-ray binaries through quasi-periodic oscillations \cite{Torok:2005}, continuum spectra of the inner edge of the accretion disk \cite{Narayan:2008}, or iron-line profiles \cite{MillerFabian:2005} indicate that a significant fraction of BHs are rapidly spinning  with $a/M \gtrsim 0.7$, with some apparently very near to maximum spin, although a few could have low spins $a/M \sim 0.1$.  However, all of three methods for spin measurement potentially suffer from model-dependent assumptions.  Moreover, these spins could have been enhanced by accretion, and may therefore not be relevant for birth spins in NS-BH or BH-BH binaries.

Population-synthesis models indicate that a majority of both NS-BH and BH-BH systems undergo a common-envelope phase with hypercritical accretion, i.e., highly super-Eddington accretion with neutrino cooling \cite{2007PhR...442...75K} (however, see \cite{Belczynski:2007}).  The details of hypercritical accretion, including the associated spin-up of the black hole, are poorly understood.  Under the assumption that all of the angular momentum at the last stable orbit is transferred to the BH, which maximizes the BH spin, O'Shaughnessy et al.~\cite{2005ApJ...632.1035O} found typical BH spins in access of $0.7$, although spins tended to get smaller for more massive BHs, which are more difficult to spin up.  However, using more conservative assumptions about accretion and spin up, Belczynski et al.~\cite{BelczynskiTaam:2008} conclude that unless a BH has a significant birth spin, it is unlikely to be spinning above $a/M \sim 0.5$.

The processes governing spin orientations are not fully understood at present.  For some isolated binaries, the spins may be preferentially aligned with each other and with the orbital angular momentum.  This will happen if all three of the following conditions are satisfied: (i) if the binary was aligned before the supernovae (e.g., through mass transfer episodes), (ii) if the supernovae kicks are small enough that the post-kick orbital plane is not too different from the pre-kick one (this is particularly likely for the more massive BHs, which are likely to have lower kicks, or for binaries that already have very short orbital periods before the second supernova \cite{Bulik:2008}, and (iii) if the asymmetric supernovae themselves don't change the spin direction relative to the spin of the progenitor star.  The conditions may be slightly relaxed for the primary, as its spin can be re-aligned with the orbital plane through accretion and the Bardeen-Peterson effect.  For dynamically formed binaries, there is no clear mechanism for preferential spin orientations, so the spin directions should be isotropic.

\section{Gravitational-wave observations and astrophysics \label{sec:astro}}

\subsection{Why do we need astrophysics to inform GW searches?}

Astrophysical inputs can significantly aid searches for gravitational waves.  Perhaps the most obvious such input has to do with the selection of the optimal configurations for advanced detectors.  Advanced LIGO will have several possibly tunable parameters, including the laser power, the reflectivity of the signal recycling mirror and particularly the detuning of the signal recycling cavity.  Choices of these parameters can dramatically change the shape of the noise spectra, as shown in the Advanced LIGO Reference Design \cite{AdvLIGOdesign}.  For example, the figure in \cite{PSD:AL} shows six different possible configurations.  Some of them trade off high-frequency sensitivity for improved sensitivity at low frequencies, or broad-band sensitivity for a lower noise floor in a narrow frequency band.  Future detectors, such as the Einstein Telescope \cite{ET}, currently have an even greater range of possible configurations.  Predictions for the rates of different classes of compact-binary coalescences and the astrophysical significance of specific searches will be very important in selecting detector configurations, particularly in the era after the first detections.

Similarly, astrophysical input can aid decisions regarding the allocation of limited human and computational resources in data analysis:  Should searches be expanded to the mass range of intermediate-mass black holes?  Given scarce resources, does it make sense to develop a new search for eccentric binaries at the expense of possible minor improvements in searches for circular binaries?  And what waveforms should be used for existing searches \cite{Abbott:2009CBCS51yr, Abbott:2009CBCS51218, Sengupta:2010}?   For instance, the relatively low masses predicted for most NS-BH and BH-BH mergers suggest that inspiral-only waveforms are sufficient for such searches, but templates that include merger and ringdown signals \cite{Buonanno:2007EOBNR, Ajith:2008} will be necessary to search for mergers involving IMBHs.  Although current searches do not include spins because existing non-physical spinning template families increase the false alarm rates unacceptably \cite{VanDenBroeck:2009}, the likelihood of significant BH spins requires spinning templates to be used for parameter estimation \cite{vanderSluys:2008b}.  And the computational difficulty of including spinning waveforms in the template banks being used for detection may be partially circumvented by ignoring those regions of the parameter space which are astrophysically unlikely, such as NS-BH binaries with very rapidly spinning ($\chi \gtrsim 0.9$) massive ($M \gtrsim 10 M_\odot$) black holes \cite{Pan:2004, 2005ApJ...632.1035O}.

\subsection{What can GW searches teach us about astrophysics?}

Once the first detections of gravitational waves are achieved, GW astronomy will allow us to utilize this new observational window to study the astrophysics of stellar and binary evolution and cluster dynamics.  

Some individual detections will already carry very exciting astrophysical information.  For instance, a single detection with an accurate BH mass measurement of $\gtrsim 100\ M_\odot$ will provide an unambiguous proof of the existence of an intermediate-mass black hole.  Any measurement indicating a significant eccentricity in a coalescing binary would indicate that dynamical effects played an important role in the binary's formation.  On the other hand, statistics from multiple observations can be compared with models to constrain astrophysical parameters.  

Perhaps the most obvious statistic is the observed rate of binary coalescences.  As discussed in \cite{OShaughnessy:2008sfr}, $N$ detections will constrain the rate of mergers to within a factor $1/\sqrt{N}$, according to Poisson statistics.  A Bayesian framework can be used to convert observed rates into additional constraints on the population-synthesis models discussed in Sec.~\ref{pop-synth}, thereby making it possible to more accurately constrain the underlying astrophysical parameters.  
For a given choice of astrophysical model parameters $\vec{\Theta}$, population synthesis codes coupled to information about galaxy distributions and detector sensitivity provide a distribution of the detectable event rate, $p(R|\vec{\Theta})$.  If an actual merger rate $\hat{R}$ is measured for a given binary type, then the likelihood that a model with parameters $\vec{\Theta}$ fits the data is  
\be
p(\hat{R}|\vec{\Theta})=\int p(R|\vec{\Theta})\ 
	\exp\left(-\frac{|\hat{R}-R|^2}{2\sigma_R^2}\right)\ dR,
\ee
where $\sigma_R$ includes both the statistical uncertainty in the measurement (equal to $\sqrt{\hat{R}}$), and the systematic uncertainty in the models.  Then the constrained posterior probability density function (PDF) of astrophysical parameters is given by the Bayes' rule,
\be
p(\vec{\Theta}|\hat{R}) = \frac{p(\hat{R}|\vec{\Theta}) p(\vec{\Theta})}{p(\hat{R})},
\ee
where the prior $p(\vec{\Theta})$ represents any a priori knowledge about the parameters, and $p(\hat{R})$ is the evidence which is set by the overall normalization requirement.

\begin{figure}[htb]
\centering
\includegraphics[keepaspectratio=true, width=0.45\textwidth]{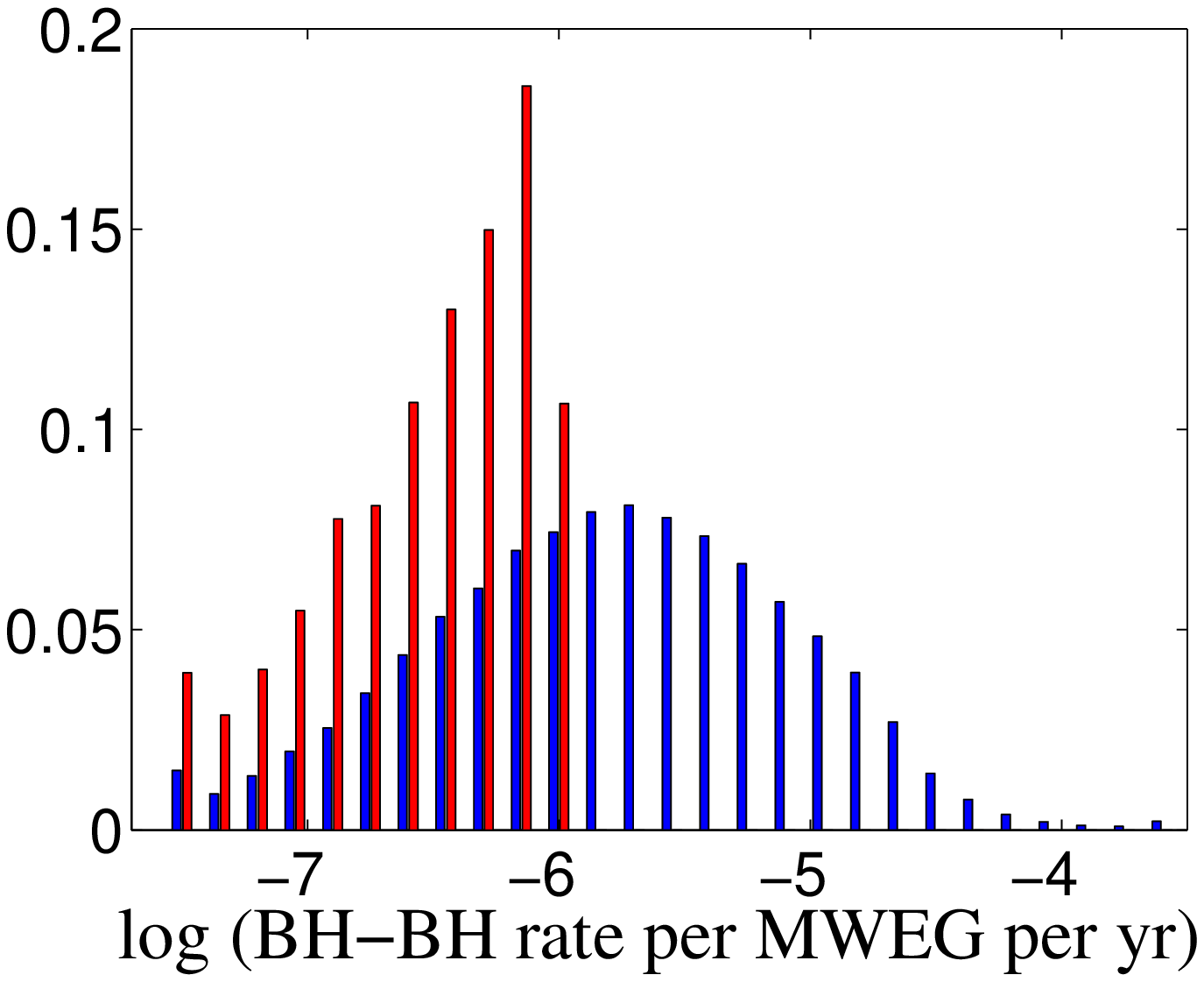}
\hskip0.25in
\includegraphics[keepaspectratio=true, width=0.45\textwidth]{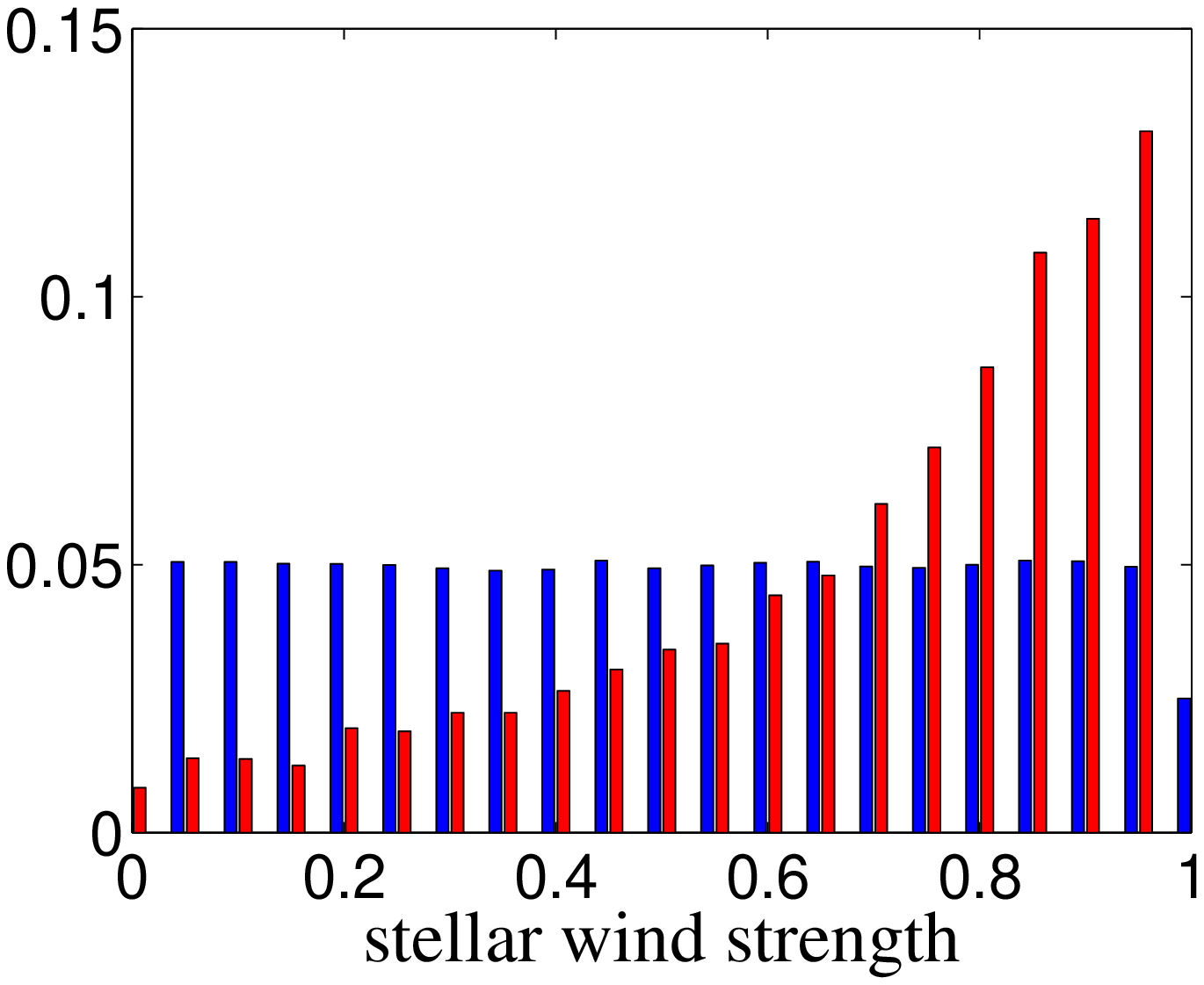}
\vskip-0.1in
\caption{The left panel shows the histogram of the BH-BH merger rate per MWEG, while the right panel shows the posterior probability distribution function of the dimensionless parameter describing the stellar-wind strength (see \cite{OShaughnessy:2005} for details).  The blue distributions result from the models analyzed in \cite{OShaughnessy:2005}, with only the observational constraint from coalescing Galactic binary pulsars imposed.  The red distributions include the additional constraint from a putative hard upper limit of 1 BH-BH merger per Myr per MWEG from GW searches.}
\label{fig:constraint}
\end{figure}

Even in the absence of detections, stringent upper limits can significantly inform our understanding of the astrophysical model parameters.  Again, a Bayesian approach can be used.  However, for illustration purposes, we can consider a simple example, where the upper limit is employed as a firm cutoff.  Consider, for example, the population-synthesis prediction for the BH-BH merger rate per Milky Way Equivalent Galaxy (MWEG), constrained by the extrapolated observations of merging binary pulsars \cite{OShaughnessy:2005}; this rate is plotted in blue in the left-hand panel of Figure \ref{fig:constraint}.  If observations placed a firm upper limit of 1 BH-BH merger per MWEG per Myr, models predicting higher rates are ruled out, resulting in the shifted distribution shown in red.  The right-hand panel of Figure \ref{fig:constraint} shows the distribution of the dimensionless parameter describing the strength of the stellar wind.  Its prior distribution is assumed to be uniform in $[0,1]$.  The blue histogram indicates that binary-pulsar observations do not constrain this astrophysical parameter, as the distribution remains flat (the drops near 0 and 1 are features of the plotting routine).  However, as the red histogram indicates, the additional constraint from the putative upper limit on the BH-BH merger rate would preferentially rule out models with low values of the stellar-wind strength, thereby significantly informing our knowledge about this astrophysical parameter.

In addition to rates, the distributions of the measured parameters of the coalescing binaries, particularly masses and spins, can be matched to the model predictions.  This approach requires several steps.  First, for each detected signal, the binary parameters have to be measured.  This involves measuring not only the best-fit parameters, but also estimating the accuracy of the measurement.  It is best accomplished by sampling the posterior PDF in the parameter space with a Markov Chain Monte Carlo \cite{Rover:2006ni, vanderSluys:2008a, vanderSluys:2008b} or an alternative Bayesian sampling technique \cite{VeitchVecchio:2008}; such techniques have been demonstrated to work for the full 15-dimensional parameter space of spinning LIGO waveforms \cite{Raymond:2009}.  Next, the PDFs from multiple detections have to be combined into a single statement about the parameter distribution in the underlying population of CO binaries \cite{Mandel:2010stat}, while taking into account selection biases, such as the mass dependence of the detection range \cite{BradyFairhurst:2008}.  At this stage, comparisons can be made to the distributions predicted by population-synthesis models.

\begin{figure}
\includegraphics{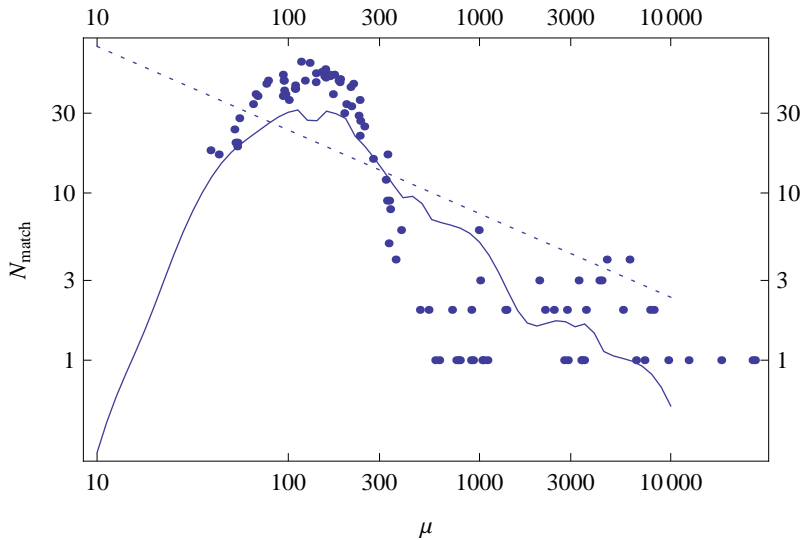}
\caption{\label{fig:MassDistributionExample}
Using the number of detections only (thin line) and the number of detections together with the observed chirp mass distribution (points) to distinguish among $238$ binary evolution models.  The horizontal axis indicates the average number $\mu$ of binary detections that each model predicts.  The vertical axis is the number $N_{match}$ of binary evolution simulations (out of the $238$ considered) whose results are statistically indistinguishable from the reference model, with 90\% confidence. For comparison, the dotted line shows $238/\sqrt{\mu}$.} 
\end{figure}

The fully Bayesian machinery described above is essential to exploit the information available in most parameter measurements, as the low-amplitude detections likely in second-generation detectors will not tightly confine their permissible ranges.  The chirp mass, however, is measured more precisely than the scatter expected between independent samples drawn from the (detected) mass distribution.  In this exceptional case the \emph{number of samples}, rather than individual measurements' accuracy, limits our ability to reconstruct the intrinsic distribution.   Assuming we do not attempt to exploit other degrees of freedom, the sampled chirp mass distribution can therefore be directly compared with the simulated distribution; see for example \cite{Bulik:2004Mc}.  A preliminary self-consistent treatment of both the number of detections and their chirp mass distribution is shown in Figure \ref{fig:MassDistributionExample}.

Finally, a wealth of information can be gained from simultaneous gravitational and electromagnetic observations of binary coalescences, such as the observations of gravitational waves associated with short GRBs \cite{LIGOGRB}.  We refer the reader to \cite{Phinney:2009, Stamatikos:2009} and references therein for detailed discussions of the scientific promise of multi-messenger astronomy, which spans the range from exposing GRB progenitors to enabling precise cosmography \cite{Nissanke:2009}.

\section{Future developments}

There is a great deal of ongoing work to improve the detection and parameter-estimation pipelines within the LIGO-Virgo collaboration.  These efforts range from the inclusion of new waveform templates and detection statistics in the pipeline \cite{Sengupta:2010} to improved sampling techniques for parameter estimation \cite{vanderSluys:2009, VeitchVecchio:2009} to simulated runs on numerical-relativity injections as part of the NINJA project \cite{NINJA}.

In order to take full advantage of the data that will soon come from gravitational-wave detectors, more work is also needed on the astrophysical side.  Larger population-synthesis runs with improved models are needed to enable comparisons against observational results.  Most simulations of dynamical-formation scenarios relevant to ground-based detectors have not yet quantitatively explored the uncertainties in the predicted rates.  In general, detailed studies of all input uncertainties in the models are needed.

We need to prepare for the era of routine detections by developing a framework for converting observed LIGO-Virgo merger rates or upper limits for compact-binary sources into improved measurements of astrophysical parameters.  In order to utilize the full wealth of information that will be available in the measured binary parameters, such as masses and spins, we must also be ready to combine the parameter PDFs from individual events into statements about the population distribution while carefully taking selection biases into account, and to compare these observed distributions against those predicted by astrophysical models.

Another direction of theoretical work relates to probing strong-field gravity, and perhaps even testing general relativity, with gravitational-wave observations.  There is a growing body of work on LISA extreme-mass-ratio inspirals as probes of the spacetimes of massive compact bodies
(see, e.g., \cite{Ryan:1995, GairLiMandel:2008, VigelandHughes:2010}).  However, it has been suggested that even ground-based detectors can be useful tools for studying possible deviations from the Kerr metric.  For example, gravitational waves from intermediate-mass-ratio inspirals (IMRIs) of NSs and BHs into IMBHs may encode enough information to allow comparisons of the mass-quadrupole moment of the IMBH against its expected Kerr value, thereby measuring the ``bumpiness'' of the black hole \cite{Brown:2007}.  Further development effort is needed to make such measurements part of the standard data analysis of detected signals.  Finally, there are ongoing efforts to study the ability of ground-based detectors to probe the equation of state of neutron stars through observations of NS-BH coalescences \cite{Hinderer:2010}.

\section*{Acknowledgments}
IM is supported by an NSF Astronomy and Astrophysics Postdoctoral Fellowship under award AST-0901985.  IM acknowledges support by NSF grant PHY-0653321 to Northwestern University for part of this work. R.O. was supported by National Science Foundation awards PHY 06-53462.   Both authors appreciate the support of the  NRDA-2 conference organizers. 

\section*{References}
\bibliographystyle{amsplain}
\bibliography{Mandel}

\end{document}